\documentclass[manuscript,nonacm,table]{acmart}

\AtBeginDocument{%
  }

\setcopyright{acmlicensed}
\copyrightyear{2024}
\acmYear{2024}
\acmDOI{XXXXXXX.XXXXXXX}

\acmConference[Conference acronym 'XX]{Conference}{June 03--05,
  2018}{Woodstock, NY}
\acmISBN{978-1-4503-XXXX-X/18/06}

\usepackage{array}
\usepackage{xspace}

\newcommand*{\credal}{\texttt{CREDAL}\xspace}

\begin{document}

\title{CREDAL: Close Reading of Data Models}

\author{George Fletcher}
\affiliation{%
 \institution{Eindhoven University of Technology}
 \city{Eindhoven}
 \country{Netherlands}}
 \email{g.h.l.fletcher@tue.nl}

\author{Olha Nahurna}
\affiliation{%
  \institution{Ukrainian Catholic University}
  \city{Lviv}
  \country{Ukraine}}
  \email{onahurna@gmail.com}

\author{Matvii Prytula }
\affiliation{%
  \institution{Ukrainian Catholic University}
  \city{Lviv}
  \country{Ukraine}}
  \email{matvii.prytula@ucu.edu.ua}

\author{Julia Stoyanovich}
\affiliation{%
  \institution{New York University}
  \city{New York}
  \country{USA}}
  \email{stoyanovich@nyu.edu}

\renewcommand{\shortauthors}{Fletcher et al.}

\begin{abstract}
Data models are necessary for the birth of data and of any data-driven system.  Indeed, every algorithm, every machine learning model, every statistical model, and every database has an underlying data model without which the system would not be usable.  Hence, data models are excellent sites for interrogating the (material, social, political, ...) conditions giving rise to a data system.  Towards this, drawing inspiration from literary criticism, we propose to closely read data models in the same spirit as we closely read literary artifacts.  Close readings of data models reconnect us with, among other things, the materiality, the genealogies, the techne, the closed nature, and the design of technical systems. 

While recognizing from literary theory that there is no one correct way to read,  it is nonetheless critical to have systematic guidance for those unfamiliar with close readings.  This is especially true for those trained in the computing and data sciences, who too often are enculturated to set aside the socio-political aspects of data work.  
A systematic methodology for reading data models currently does not exist.  
To fill this gap, we present the \credal methodology for close readings of data models.  We detail our iterative development process and present results of a qualitative evaluation of \credal demonstrating its usability, usefulness, and effectiveness in the critical study of data. 

\end{abstract}

\maketitle

\section{Introduction}
\label{sec:intro}

As the saying goes, knowledge is power.  Through knowledge, communities make their worlds.  Yet, as has been highlighted since Foucault, the converse also holds: power is knowledge \cite{foucault}. It is primarily those with power who are enabled to resource the data systems that drive contemporary data-driven decision and knowledge making.  This, in turn, leads to a perpetuation of the status quo, where what is known is predominantly for and by those with power.  There is a vicious cycle in data systems of power begetting more power.  Consequently, data systems are too often key enablers and amplifiers of the cruelties of the status quo \cite{ansorge,becker,benjamin,bode,costanza,couldry,eubanks,stevens,LewinskiBS24,biopower,weizenbaum}.

\begin{figure}
    \centering
    \includegraphics[width=0.5\linewidth]{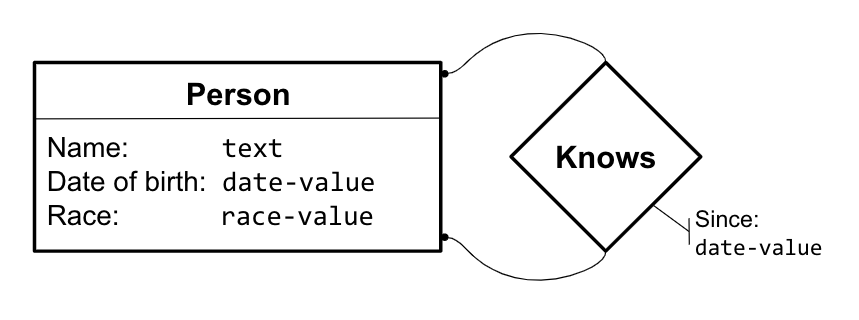}
    \caption{Data model for Example~\ref{ex:1}.
   In this data model, \texttt{Person} is a class and \texttt{Knows} is a relationship between elements of this class; furthermore, instances of class \texttt{Person} have ``name'', ``dob'',  and ``race''  attributes, and \texttt{Knows} relationships have a ``since'' attribute. 
    As an example data instance of this model: Saori is a \texttt{Person} (name ``Saori'', race ``Asian'', and date of birth 2001) who \texttt{Knows} Kotaro (who is also a \texttt{Person}, name ``Kotaro'', race ``Native Hawaiian or Other Pacific Islander'', and date of birth 2002) since 2018.    }
    \label{fig:example}
\end{figure}

In this work, we highlight an emerging thread in the broader conversation of how to open up and intervene in cycles of data for more equitable and just data futures.  We focus on a seemingly mundane yet critically vital aspect of data and information system design and engineering: data modeling.\footnote{Here, a data model is also known as a conceptual model, an ontology, a knowledge graph, an entity-relationship diagram, a T-box, a set of features, a set of variables of an experiment, a semantic network; basically, a collection of classes and relationships between these classes. See Figure \ref{fig:example} for an example.}  Every information system (every algorithm, every machine learning model, every statistical model, every AI solution, every database) has an underlying data model without which the system would not be usable.  Data models, whether implicit or explicit, are present at the birth of data and are necessary for the design, development, and use of any data system. Hence, data modeling is a core topic taught in computer science, data science, software engineering, information systems, and information science degree programs.

\begin{example}
\label{ex:1}
Consider the data model visualized in Figure \ref{fig:example}. 
Here, ``race'' takes a single value drawn from a fixed list, standardized ``for all Federal reporting purposes'' in the USA: American Indian or Alaska Native, Asian, Black or African American, Native Hawaiian or Other Pacific Islander, and White.\footnote{\url{https://www.doi.gov/pmb/eeo/directives/race-data}}
    What does it mean that race is single-valued and drawn from a fixed domain?  In which worlds is this meaningful?  How does it conflict with other worlds, where race is multi-valued and open-ended?  To what ends does the data system even model race in the first place?  Why is race an attribute of people, i.e., modeled as inherent (objectively) in a person?  Or are people racialized by other people, in which case race would better be modeled as a separate class or as a relationship between people?  In what world is race a fixed, limited, single, unmovable, inherent feature of people?  Why is the data system provider perpetuating this world?~\footnote{See 
    Chapter 6: The case of race classification and reclassification under apartheid, 
    Bowker and Star~\cite{sorting}. }
\end{example}

How can we systematically unpack a data model, to interpret it as the contextualized creative cultural artifact that it is?  Drawing inspiration from literary criticism,  our proposal is to closely read data models in the same spirit we read literature.   Close reading is a well-known approach to literary analysis that involves careful and sustained interpretation of textual artifacts \cite{frank}. Questions to consider during close reading include the speaker, the audiences (intended or otherwise), the reader's assumed knowledge, the purpose of included details, ambiguous words or phrases, patterns, rhythm, movement, and anything that raises questions or requires clarification.  

Students are often taught step-by-step approaches for conducting a close reading of a literary passage.\footnote{e.g., \href{https://writing.wisc.edu/handbook/closereading/}{U. Wisconsin Madison Writer's Handbook}, \href{https://www.york.ac.uk/english/about/writing-at-york/writing-resources/close-reading/}{U. York -- Writing at York}}
This typically begins with selecting a short section of text and paying attention to unusual or repetitive imagery and themes. The reader is then guided to read the passage, taking notes on language use, repetitions, and changes in tone. The analysis phase involves examining elements such as diction, narrative voice, and rhetorical devices. Following this, a descriptive thesis is formulated based on language observations, and finally, an argument is developed, connecting language use to the broader themes of the text. The method emphasizes the importance of understanding not just how language is used but why, encouraging a thoughtful exploration of the author's intentions and the text's overall significance.

\subsection{Why close readings of data models?}
While working with (and within) data models, we bracket off the worlds behind the model's abstraction, often to the point that we lose contact. This bracketing is, after all, the purpose of modeling: to highlight some descriptions of the world over other descriptions.  For many doing the technical work in computing and data science, this contact is never made (or held in abeyance, as ``low value'' messy labor, the dirty wrangling work of data  \cite{SambasivanKHAPA21}).  

Close readings of data models reconnect us with, among other things:

\paragraph{The materiality of data models}  
As Brian Cantwell Smith points out in early work on the limits of modeling in information systems,   ``one of the most important facts about computers ... is that we plug them in'' \cite{Smith85}.  This highlighting of the materiality of computing was taken up by Paul Dourish's study of data systems,  underscoring that ``the material arrangements of information ... matters significantly for our experience of information and information systems'' \cite{dourish}.  Too often, data is discussed as though it were immaterial, abstractions floating above the physical world. Close readings of data models can draw our attention to the grounding of data in the physical world and infrastructures, with physical effects and side effects in our bodies, embodied communities, and environments \cite{hogan}:
{\em  Where does this data model (and its data) reside?  Is it hosted in the cloud?  Where are the data centers for the cloud provider?  What are the reasons for this and the implications (organizational, legal, social, political, environmental, ethical, ...)? ...}

\paragraph{The closure of data models}  
Sabina Leonelli's study of data journeys has underscored the relational nature of data, namely, that data exists only in relation to a particular context, narrative-making community, and moment of inquiry
\cite{datajourney,Leonelli2015}.
As such, a data model is an interpretation, a biased window onto life \cite{feinberg,maleve}, a closure, a forgetting of the limitless alternatives, the unbounded space of interpretations not taken \cite{parrish,onuoha,azoulay,sherman}.  
And the information systems built around the model and their deployment live within the closure,  enabling and enacting the world of the closure, for better and for worse \cite{ansorge,bopp,martin,LewinskiBS24}.
As a rich example, von Lewinski et al.\ have carefully demonstrated this worlding power of data models in the legal domain \cite{LewinskiBS24}.
Close readings help us to open the closure, to find the seams and stitches that hold it together, and to surface what has been left out, and why, and what has been enclosed, and why:
{\em Why do we have this class?  Why these attributes?  Who views the world this way?  Who doesn't?  What does it mean to view the world this way?  Who benefits?  Who is left out? ...}

\paragraph{The genealogies of data models} 
Technologies such as data, data models, and data systems have histories, histories which we inhabit as we work and live with and within concrete data models.  For example, recent scholarship has continued to deepen our understanding of how datafication, data analytics, and data systems have been intertwined with the imperatives of slavery, eugenics, war, colonialism, and state and industrial surveillance and control of citizens and workers (e.g.,  \cite{penn,valdivia,whittaker,wiggins}). 
Scholarship in programming languages and software engineering is also highlighting the genealogies of these technologies (e.g., \cite{parrish,hermans}).   Close readings help us to connect the data model at hand to the broader worldviews and ideologies of datafication: 
{\em Why is this a data problem in the first place?  What are alternatives to datafying this problem?  Why is this a problem at all?  What are the implications of problematizing this activity/inquiry?  What are the implications of taking non-data-centric approaches?  Are we constraining future possibilities by taking a data-driven approach now? Which futures does this data model imagine? ...}

\paragraph{The techne of data models} Close readings can help us to unpack and understand the production and engineering of the model, namely, the tools, methods, and practices employed during the life of the model:  
{\em Which software tools were used to create the model?  Was the model influenced by this tooling?  Is a data system or platform used to implement the model?  If so, what are the modeling functionalities of the system (e.g., schema language, constraint language, logical modeling language)?  Are these functionalities and capabilities suitable for the data model?  If not, what are the workarounds?  Is there an expected or known query or analytics workload?  Can the system support this workload? Will the data be governed?  What governance infrastructure is available? ...}

\paragraph{The design of data models} Shining light on the materiality, genealogy, closure, and techne of data models, close readings help us to shift registers, from the classical textbook view of modeling as a description and documentation of a given fixed reality, to a view of modeling as design, as a creative socio-political activity of diplomacy, compromise, power struggles, and consensus making \cite{Feinberg17,shaw,simsion}.  Close readings lead us to ask: {\em Who was involved in making this model?  What are their roles?  What are their professional, organizational, and personal relationships?  Who led the process?  Who initiated the process?  Was the process hindered -- technically, organizationally, legally, politically, ...? Are there open research challenges here to remove (or introduce) friction or facilitate (or inhibit) design activities? ...}

\subsection{The challenge: How to read data models?}
To our knowledge, a systematic methodology for reading data models currently does not exist.  While recognizing from literary theory that there is no one correct way to read \cite{frank}, and that in fact every encounter of a reader with a text is a non-repeatable experiment, it is nonetheless critical to have a starting point for those unfamiliar with close readings.  This is especially true for those trained in the computing and data sciences, who too often are enculturated to set aside the socio-political aspects of data work.

\subsection{Goals and Contributions}
In our work, we have developed \credal, a structured methodology for the close reading of data models.  Such a methodology must be designed to facilitate a nuanced exploration of language, relationships, and patterns within the data models, aiming to uncover the biases within modeling choices and contribute to the creation of fair and just knowledge representations.
The goal of this research extended beyond the mere development of \credal; it was equally focused on ensuring its effectiveness, usefulness, and applicability in practice. Therefore, we guided our research using the following research questions:

\begin{enumerate}
    \item [\textbf{RQ1:}] Is \credal usable and useful? 
    \begin{enumerate}
        \item [\textbf{RQ1.1:}] Is \credal easy to learn and apply?
        \item [\textbf{RQ1.2:}] Does \credal improve data modeling proficiency?
        \item [\textbf{RQ1.3:}] Are learners likely to use \credal in the future?
    \end{enumerate}
    \item [\textbf{RQ2:}] Does \credal help understand, design and critically evaluate data models?

        \begin{enumerate}
            \item [\textbf{RQ2.1:}] Does \credal help with understanding data models?
            \item [\textbf{RQ2.2:}] Does \credal alter the approach to structuring and modeling data within modeling tasks?
            \item [\textbf{RQ2.3:}] Do learners experience a change 
            in their data modeling perspectives after working with \credal?
        \end{enumerate}
\end{enumerate}

We will present the results of a qualitative study towards answering each of these questions,  highlighting the \emph{perceived benefits} of \credal, and providing concrete indications for further research and development of the methodology.   
\section{Background and Related Work}
\label{sec:background}

In the previous section, we have highlighted data models as imbricated in making possible and driving the social, material, and genealogical facets of the work of data.  We also positioned close readings of data models with respect to the classical notion of close reading of texts.  In this section, we place our work in the context of closely related work in reading and modeling data.

\subsection{Reading data}

We first highlight work on reading data and data models.
Lindsay Poirier identifies three modes of \emph{close reading for datasets}: denotative (understanding technical aspects), connotative (understanding cultural contexts of production and change), and deconstructive (understanding representational limits) \cite{poirier}.
Through case studies, Poirier's work shows how students learn to read datasets through these modes to better understand the assumptions and politics of a given dataset.  

Melanie Feinberg has proposed reading data models as a ``slow'' data interaction paradigm which is complementary to the dominant ``fast'' outcome-oriented search and retrieval paradigms \cite{melanieSlow}.   Slow readings are shown to create opportunities for teasing out and reflecting on the interpretative facets of data model interaction.
While highlighting the vital importance of reading data, the work of both Poirier and Feinberg does not introduce a systematic methodology to jumpstart and guide close readings, which is important for beginning readers (especially those from STEM backgrounds). 

Karen Wickett and Nikki Lane Stevens have recently introduced \emph{critical data modeling} as a focus within critical information and data studies.
Wickett's work \cite{wickett} identifies a basic representation model to highlight the propositional, symbolic, and material aspects of data models. This model aims to support \emph{close readings of data collections} through the lens of these aspects as well as the transformations between the modeling domain and models of the domain in information systems.  The goal of this work is to elevate models as first-class citizens in the field of critical data studies, thereby making the biases and assumptions latent in the models under which data collections are created visible and open to critique and contestation.  A notable aspect of Wickett's model is 
the focus on the propositional content of data collection---on the truth claims of data relative to a posited ``real world''---rather than the relational aspects of data collections.  Here, people and modeling (and iterative remodeling) activities are secondary; both the modelers themselves and those being modeled are essentially absent.  This is in contrast with a relational understanding of data and data models, which highlights the hermeneutic nature of data-driven knowledge making (i.e., data and data models are only ever such in relation to human storytelling).

Lane Stevens also aims to elevate data models as first class citizens in the critical study of data, through several autoethnographic investigations that uncover the fundamental ways in which data models make algorithmic decision and knowledge-making possible~\cite{stevens}. Lane Stevens particularly unpacks the ways in which modeling choices are intimately tied up in social power and empowerment.  This work demonstrates how decisions and knowledge embody and perpetuate the social norms and power structures coded in data models.  The overarching contribution of this work is to identify data models as sites for intervention and subversion towards more just and equitable futures.  

We also highlight qualitative studies of troubling data models with similar aims to ours, including grounded theory investigations of Michael Muller et al. on how data scientists work \cite{muller}, and the recent ethnographic ``autospeculation'' method of Brian Kinnee et al. \cite{kinnee}.  
Finally, we note the broad literature on \emph{critical data literacy}, which aims in part to help students understand the complex social factors behind data collections  
\cite{dangol}.  Our work is complementary with several of the aims of this area of scholarship.

\subsection{Modeling data}

Conceptual and knowledge modeling of data has been studied in the data management, business information systems, and knowledge representation communities for over 50 years.  For overviews of these rich literatures, see the surveys of Veda Storey et al. \cite{akoka,storey} and Wei Yun et al. \cite{yun2021}.  Apart from a rich sub-literature on model validation (i.e., tools and methods to ensure that a model is fit for purpose), we find that the overwhelming majority of work has overlooked broader social aspects of data modeling.   Notable exceptions include the qualitative investigation of Graeme Simsion and colleagues that highlights the constructive consensus-making nature of modeling in business information systems \cite{simsion}, and recent efforts in the conceptual modeling community on inclusiveness in modeling, see  Lukyanenko et al. \cite{LukyanenkoBSP023}.   
\section{Developing the Close Reading Methodology (\credal)}
\label{sec:credal}

This section outlines the iterative process we followed to develop \credal. The process evolved through a series of steps, each designed to refine the methodology based on practical applications, feedback, and structured evaluation. Below, we detail each step of this process.

\subsection{Initial development}
\label{sec:credal:dev}

\subsubsection{Initial Exploration and Refinement}
\label{sec:credal:dev:1}

Our research team initiated the process by deeply engaging with the technique of close reading, traditionally employed in literary studies for detailed analysis of texts~\cite{frank}. Following a review of the relevant literature and practical application on literary texts, we sought to adapt this approach to the analysis of data models. Each team member independently proposed methods for applying close reading to the schemas. After reviewing individual work, we compared our approaches, identified commonalities, and developed the first draft of a methodology for applying close reading to data models.

In the next phase, we applied the newly developed methodology to a fresh set of schemas. This time, all team members used the same standardized approach, allowing us to observe how different individuals interpreted and applied the same methodology in diverse ways. This approach highlighted variations in interpretation and application, which led us to further refine and modify the methodology.

We used both knowledge graph data models (examining schemas from Wikidata.org and Schema.org) and relational data models during this first step. 
Recognizing the widespread use of relational data models \cite{silberschatz} across various industries and the broad availability of standard relational data modeling curricula, we decided to focus on relational data models for the remainder of our study.
We conducted another round of close reading, this time using open-source 
data models,\footnote{e.g., \href{https://gitmind.com/erd-examples.html}{GitMind}, \href{https://devtoolsdaily.medium.com/crafting-an-automatic-erd-generator-a-journey-from-ddl-to-diagram-83cc5da8cab7}{DevTools Daily}, \href{https://www.conceptdraw.com/How-To-Guide/entity-relationship-diagrams}{ConceptDraw}}
and adapted our methodology based on this experience.

\subsubsection{Development of Supplemental Materials}
\label{sec:credal:dev:2}

In order to verify the applicability and effectiveness of the methodology, it was necessary to involve more independent reviewers.
Therefore, to ensure that the methodology could be effectively applied by others, we developed a set of supplemental materials. Since different team members brought unique perspectives and strategies to the project, we synthesized these insights into a practical guide. This guide contains valuable tips and advice aimed at helping users navigate potential challenges during applications of the methodology.

Additionally, we created a sample reading to demonstrate the methodology’s practical application. The chosen example, ``A Secure Students’ Attendance Monitoring System''~\cite{erd} (see Appendix~\ref{sec:example_reading} for details), was specifically selected for its balanced complexity, ensuring that it was detailed enough to be informative without overwhelming learners. The familiar context of education made the example relatable to a wide audience, and the step-by-step approach offered clear guidance. Together, these supplemental materials served as essential resources, providing both conceptual support and practical reference points.

\subsection{Iterative refinement}
\label{sec:credal:refine}

We needed a systematic approach to gathering deeper feedback on \credal to iteratively refine the methodology. Moreover, systematic feedback was necessary to answer the research questions guiding our study. Consequently, we developed a set of semi-structured interview questions to gather feedback from individuals working with data models.  We engaged with three groups of people, of increasing size, first explaining the \credal methodology to them and then collecting their feedback.

\subsubsection{Initial Interviews with Undergraduate Students}
\label{sec:credal:refine:1}

Each team member recruited a volunteer Computer Science student as a reviewer. The students were provided unlimited time to familiarize themselves with \credal and its supplemental materials. They were also given unlimited time to apply the methodology to a relational data model, which was provided to them. The application of \credal was carried out in an unconstrained written form according to the student's preferences.

Once the students reported that they had completed the application of the methodology, they were invited to participate in a 30-minute semi-structured interview. These interviews were recorded using an audio recorder, and the anonymized recordings were subsequently transcribed manually by team members. The text-based transcriptions were then used for further analysis, enabling us to gather detailed feedback. Based on this feedback, we identified several areas for improvement and addressed key gaps in the methodology, enhancing its clarity and making it more intuitive and accessible for users.

\subsubsection{Organizing a Workshop for Peer Feedback}
\label{sec:credal:refine:2}

After refining the methodology, we conducted a workshop with 6 graduate students in computer science and data science at New York University. Prior to the workshop, participants were invited to review the \credal methodology and its supplemental materials to familiarize themselves with the process. The workshop itself was one hour long, during which participants collectively applied the methodology to the same relational data model as the previous respondents. Following the application of the methodology, feedback was gathered using the same set of interview questions.

Using the same data model and identical interview questions ensured consistency and allowed for a comparison of different versions of the methodology. This approach also minimized variability in the feedback that might have arisen due to differences in the size, complexity, or format of the data models. Additionally, using the same interview questions enabled us to directly track the progress of \credal's development and evaluate the effectiveness of the improvements we made.

From the feedback collected during the workshop, we gained valuable insights into how to make \credal easier to understand and use.
One key recommendation was to increase the use of visual aids to accompany the methodology, to help participants familiarize themselves with the methodology more quickly.

\subsubsection{Conducting a Pilot Study}  
\label{sec:credal:refine:3}

To address these recommendations, the next group of respondents, comprising 4 graduate students in data science and 7 undergraduate students in computer science, were provided with a video guide. This guide presented the key points of the methodology in a visual format, with examples illustrating its application. Respondents were also given an improved version of the methodology guide, which included practical tips to facilitate the application process, along with a completed example of applying \credal to a data model.  We will provide additional details about the pilot study in Section~\ref{sec:eval}, and will discuss results of evaluation in Section~\ref{sec:discussion}.
  
\section{A Structured Guide to \credal}
\label{sec:credal:guide}

In this section, we present a structured guide to \credal (Section~\ref{sec:credal:guide:steps}), along with other materials we developed to support the adoption and use of the methodology (Section~\ref{sec:credal:supp}).  

\subsection{\credal}
\label{sec:credal:guide:steps}

\begin{enumerate}
\item \emph{Define Research Goals and Understand the Data Model.} Set clear objectives for your analysis, whether it be model-driven or, if data is available,  data-driven.
Understand your data and data model, paying attention to its structure, domain, and other details, such as missing relevant details.

\item \emph{Evaluate the Context, Domain Knowledge and Sources.} 
(a) Gain domain-specific insights from additional sources. 
(b) Consider ethical, privacy, and other concerns, including the absence of sensitive but relevant data. 
(c) Evaluate data sources for their biases and omissions.

\item \emph{Exploratory Data Analysis (EDA) and Schema Analysis.}
\begin{itemize}
    \item For data-driven close reading: Identify general patterns in the data, including outliers and unexpected features (for more practical tips on EDA, see \cite{abedjan,downey}). 
    \item For model-driven close reading: Conduct a detailed schema analysis, focusing on the following aspects:
        (a) Identifying entities and relationships;
        (b) Reviewing constraints such as primary keys, foreign keys, unique constraints, and field formats; and,
        (c) Identifying indices and design patterns.
\end{itemize}

\item \emph{Related Schemas Exploration.}
    (a) Identify related schemas and assess their content in comparison to the primary schema;
    (b) Note any elements or relationships that are included or excluded in the related schemas; and,
    (c) Analyze how the comparison can inform improvements or enhancements to the target schema.

\item \emph{Assumptions Loop.}
\begin{enumerate}
    \item Select a small portion of the data model (e.g., an entity, its attribute, or a relationship between entities) where potential interesting bias may be present.
    \item Establish criteria for determining fairness and objectivity before identifying negative bias or inequality. It is essential to distinguish between assumptions and verifiable bias.
    \begin{itemize}
        \item For data-driven close reading: Define metrics to quantify assumed bias or inequality. These may include skewness in data distribution, the presence of empty fields, or other quantifiable characteristics.
        \item For model-driven close reading: Compare the schema with related models, analyze similar fields, and review restrictive types or irregular relationships to validate assumptions. Conduct thorough domain research to substantiate any findings.
    \end{itemize}
    \item Identify potential risks or issues arising from the assumed bias. Consider specific scenarios in which the model may fail or produce unintended consequences.
        i) Assess how sensitive attributes (e.g., gender) influence decision boundaries and overall fairness when legally and ethically permissible;
        ii) Evaluate fairness by analyzing causal pathways and understanding how different factors contribute to model outcomes; and
        iii) The "5 Whys" method is recommended to trace the root cause of bias and understand its origin (details in Section~\ref{sec:credal:supp}). 
    \item Propose solutions to mitigate identified harmful biases and/or enhance the comprehensiveness of the data or schema.
\end{enumerate}

\item \emph{Compile Findings.}
    Generate a conclusion summarizing your observations on the analyzed model. This report should be easily comprehensible for the target audience and ideally provide valuable insights to aid the model developer in its revision.
\end{enumerate}

\begin{figure}[t]
  \centering
  \includegraphics[width=0.9\linewidth]{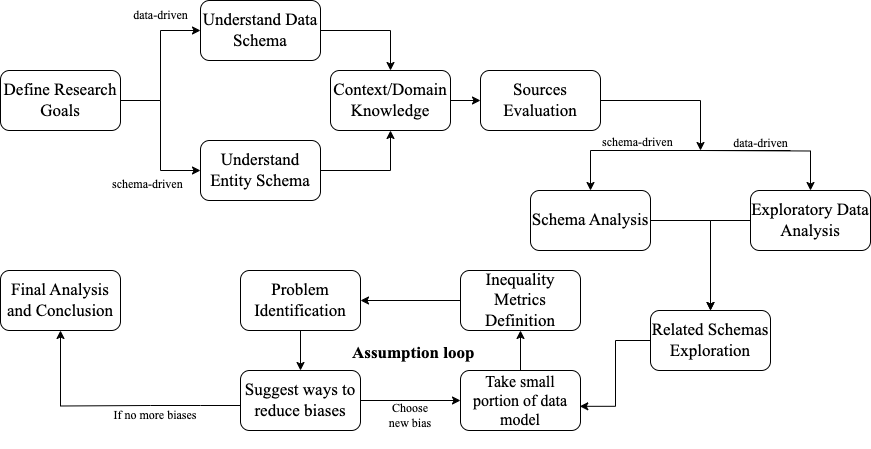}
  \caption{Visual Representation of \credal}
  \Description{Directed graph depicting the main steps to take during data model close reading}
  \label{fig:close_reading_pipeline}
\end{figure}

The overall pipeline of the data close reading methodology is provided in Figure~\ref{fig:close_reading_pipeline}.

\subsection{Supporting Materials}
\label{sec:credal:supp}

\paragraph{Video tutorial.} As a result of the iterative improvements to the methodology, we identified the need to enhance the visual presentation of the methodology to improve comprehension. In response, we developed a video tutorial that presents and explains the key terminology and main stages of the methodology, reinforcing each step with illustrative examples. 

\paragraph{Example of Reading.} To demonstrate the application of the methodology, we developed a comprehensive example reading that describes the entire process from start to finish, illustrating the nuances involved in applying the methodology to a real-world scenario. The selected data model, ``A Secure Students’ Attendance Monitoring System'' (Appendix \ref{sec:example_reading}), was chosen due to a reasonable level of complexity. It provides sufficient depth to thoroughly illustrate \credal while remaining accessible to the target audience. This step-by-step walk-through offered clear instructions, making the methodology application more straightforward for users.

\paragraph{Suggestions and Best Practices.} We identified several key techniques that enhance both the perception and application of the methodology. These insights are organized into practical recommendations:

\begin{itemize} 
    \item \textbf{Use the ``5 Whys'' Method.} This technique can help identify bias and uncover its root causes. The process involves asking ``why'' multiple times to drill down into the underlying issues. The steps are as follows:
        \begin{itemize} 
            \item Define the problem: Clearly articulate the issue to ensure an accurate understanding, as solving an ill-defined problem is challenging. 
            \item Ask the first ``Why''. Begin by questioning why the problem occurred to uncover initial insights. 
            \item Continue asking ``Why''. For each identified reason, continue asking ``why'' until you reach the fundamental cause, repeating this process until a clear solution emerges. 
            \item Practical Tips: Move swiftly from one ``why'' to the next to maintain focus and avoid distractions. Know when to stop, as continuing beyond meaningful responses may lead to diminishing returns. 
    \end{itemize}
    
    \item \textbf{Incorporate Brainstorming Sessions.}  Brainstorming encourages creative thinking and diverse viewpoints, which helps uncover biases that may not be evident in a solo analysis. Key steps include:
        \begin{itemize}
            \item Visualize the goal of the session.
            \item Document all discussions and ideas.
            \item Encourage participants to think aloud and propose varied ideas.
            \item Foster an environment where all ideas are welcomed without immediate criticism.
            \item Collaborate and ask clarifying questions.
            \item Organize the outcomes of the brainstorming session for further analysis.
        \end{itemize}
        
    \item \textbf{Apply the Principles of Literary Close Reading.}  Close reading, rooted in literary analysis, offers a systematic approach to examining textual content in depth. Its principles can be adapted for data model reading:
        \begin{itemize}
            \item Focus on explicit content, avoiding speculative interpretations.
            \item Read slowly and attentively to capture nuanced details that might otherwise be overlooked.
            \item Perform multiple readings, allowing for a comprehensive understanding of the material.
            \item Analyze each element thoroughly to ensure a detailed and accurate interpretation.
        \end{itemize}
    \end{itemize}
    
    \paragraph{Common Pitfalls and Considerations.} When applying the methodology, users may encounter several potential challenges, summarized below, along with advice for mitigating them:
    \begin{itemize} 
        \item \textbf{Overlooking Entity Relationships.}
        Thoroughly explore the relationships between entities within the schema. Neglecting these connections may lead to incomplete or inaccurate conclusions.
        \item \textbf{Managing Large Data Volumes.}  
        A data-driven approach may introduce complexity when dealing with large datasets. Focus on specific areas of interest and relevant data points to avoid becoming overwhelmed by the volume of information.
        \item \textbf{Bias Below the Surface.}  
        Bias is not always apparent at the top levels of the schema. Take multiple perspectives, playing the roles of both the creator and reviewer, to identify potential biases or controversies that may lie deeper in the structure.
        \item \textbf{Working with Unfamiliar Data or Schemas.}  
        When dealing with an unfamiliar domain, consider consulting specialists or leveraging external resources to gain a better understanding before proceeding with bias analysis.
        \item \textbf{Curiosity and Questioning.}  
        Embrace curiosity and do not hesitate to ask fundamental questions. Seemingly basic inquiries can often lead to important discoveries and unveil hidden aspects of the data model.
        \item \textbf{Concluding the Analysis.}  
        After identifying biases, it is essential to clearly articulate your findings and suggest actionable solutions. This ensures that the issues are addressed and mitigated in future iterations of the data model.

    \end{itemize}

\section{Validating CREDAL}
\label{sec:eval}

Incorporating human participants into the evaluation process provides valuable insights into the methodology's ability to achieve our intended objectives. This allows for an evaluation of whether the methodology effectively facilitates a deeper critical understanding of data models. Additionally, engaging a larger audience in the testing phase helps mitigate biases by capturing a range of perspectives, revealing potential ambiguities in seemingly straightforward concepts, and clarifying those that may initially appear complex. This process aids in validating both the content and the structure of the methodology, determining if it is accessible, appropriately concise, or comprehensive for the subject matter, and identifying other aspects that can only be revealed through application and feedback.

\paragraph{Audience selection.} To validate \credal, we conducted interviews with 11 students from applied science programs. Specifically, we combined two academic groups, interviewing 4 students from an MS program in data science and 7 students from a BS program in computer science at Ukrainian Catholic University in Lviv, Ukraine.\footnote{Our study received approval from the institution's ethics review board.}  Considering that \credal is intended for technical students and practitioners involved in data modeling and related tasks, including dataset creation, data analysis, and database design, this selection allows us to assess the relevance and utility of the methodology for its primary target audience.

\subsection{Interview process}
\label{sec:eval:process}

We started by conducting a workshop with participants to introduce the \credal methodology with the help of materials described in Section~\ref{sec:credal:guide}.  We then gathered their feedback on \credal using a semi-structured interview protocol.  The basic interview structure consisted of 14 questions, listed in Appendix~\ref{sec:questions}.  We supplemented this structure with follow-up prompts when necessary, particularly when clarification was needed.  Questions were divided into four categories:

\begin{itemize}    
    \item Participant data modeling background and experience, and their experience with \credal.  General feedback on the methodology, including its strengths and weaknesses, perceived change in own understanding of data modeling, and confidence in own data modeling skills. 

    \item Feedback on supporting materials, including a video tutorial and close reading example.  Feedback on the usefulness of the literary close reading analogy for learning \credal.

    \item Perceived effectiveness of \credal, including its practical applicability and the likelihood that participant would use the methodology in their work or studies.
        
    \item Feedback on how \credal may be improved in the future.
\end{itemize}

\subsection{Interview coding}
\label{sec:eval:coding}

We recorded and transcribed the interviews and then coded them using the codebook presented in Appendix~\ref{sec:Codebook}.  This codebook was created manually, based on consensus among researchers over multiple rounds of independent coding and follow-up discussions.  We also used Atlas.ti to help organize interview data (i.e., associate the manually generated codes with quotes in the interview transcripts). 
Note that Atlas.ti offers the capability to use Generative AI to generate relevant codes, including the option to align these codes with specific research questions for enhanced contextual relevance.  However, we opted for manual coding in this study. Given the novelty of our research thesis, manual coding ensured that no critical details were overlooked and allowed for more accurate identification of recurring themes.

To mitigate bias, we independently tagged interview transcripts using two approaches, \emph{questions-based} and \emph{context-based}.
\begin{itemize}
    \item[1.] \textbf{Questions-based coding method}:
    To develop a set of codes, we first reviewed the interview questions and established corresponding code groups. Based on participants' responses, we identified the most frequently mentioned themes, created relevant codes, and assigned them to the predefined code groups. With this approach, we generated 8 groups with 31 codes.

    \item[2.] \textbf{Context-based coding method:}
    Under this approach, we began by analyzing all interviews to identify the most frequently occurring topics. Codes were created for these phrases based on the context, and using these codes, we derived the names for the respective code groups. The codes were fine-grained, making them more specific and less repetitive. With this approach, we generated 5 groups with 63 codes.
\end{itemize}

After using these two approaches, we decided to analyze their similarities and differences. This process was also carried out manually, with the authors of the two previous approaches looking for corresponding codes that were assigned to the same quotation (marked place in the interview text) or had identical meanings.\\

In the case of code groups, we decided to keep four groups that best corresponded to our objectives and to the research questions:
\begin{itemize}
    \item[A:] Participant background (3 codes)
    \item[B:] Methodology effectiveness (3 codes), relevant for research questions \textbf{RQ1.2}, \textbf{RQ1.3}, \textbf{RQ2.1}, \textbf{RQ2.2}, and \textbf{RQ2.3}
    \item[C:] Methodology improvement (6 codes)
    \item[D:] Methodology strengths (3 codes), relevant for research question \textbf{RQ1.1}
\end{itemize}

All codes, code groups, and their explanation can be found in Appendix~\ref{sec:Codebook}.   
\section{Results analysis and discussion}
\label{sec:discussion}

After conducting and analyzing all 11 interviews, we found that each participant had prior experience with data modeling, with 2 reporting basic data modeling proficiency, 7 reporting intermediate proficiency, and 2 reporting advanced proficiency. Further, none of the participants were familiar with the literary close reading methodology prior to the workshop.

\begin{figure}
    \centering
    \includegraphics[width=0.7\linewidth]{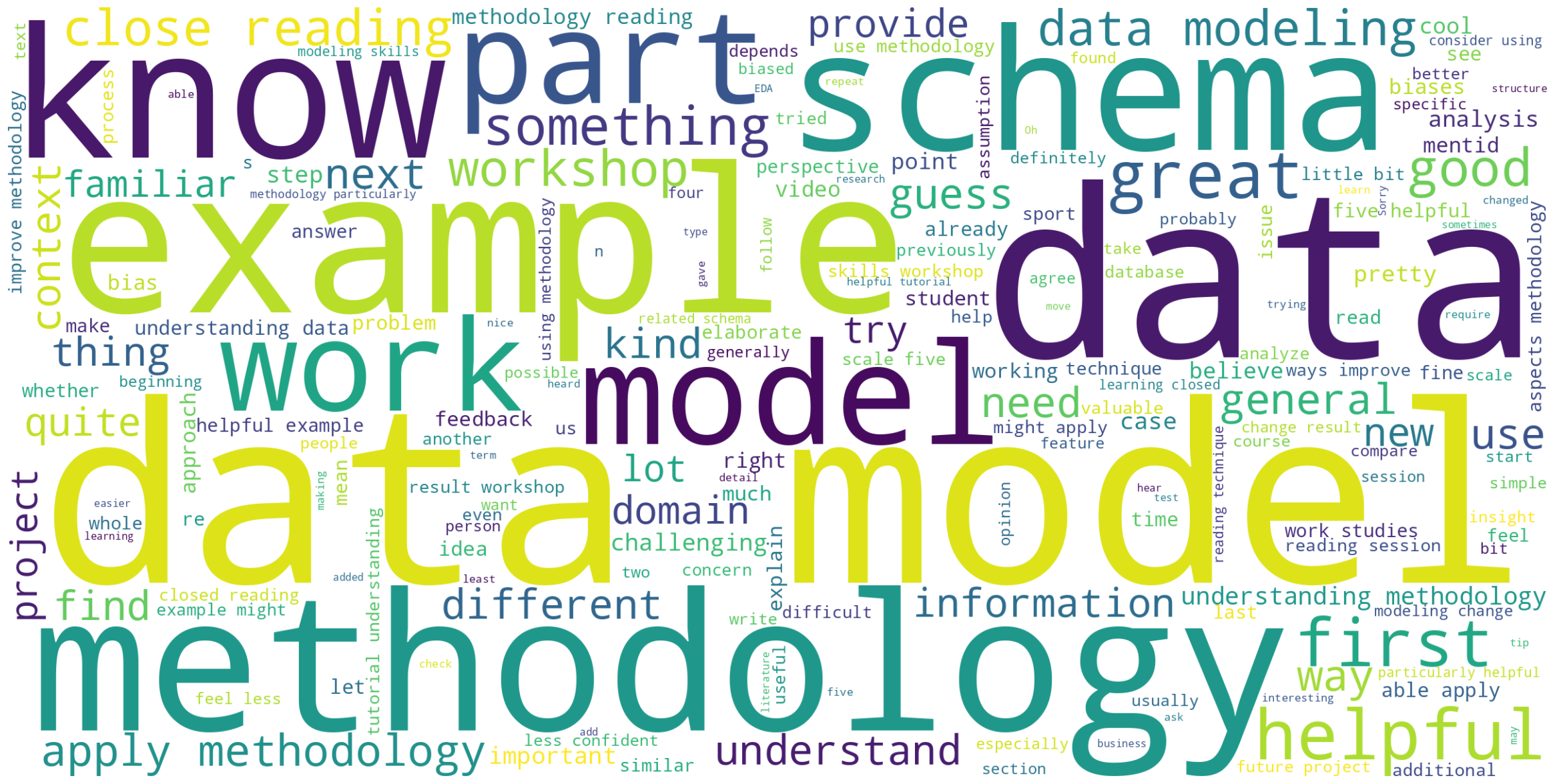}
    \caption{Word cloud of all participant responses during the interviews}
    \label{fig:word_cloud}
\end{figure}

While each participant expressed unique thoughts and opinions, there were also significant commonalities among them. By synthesizing the insights from all interviews, we identified notable patterns, which we will discuss throughout this section.  To start, we present the results of a simple quantitative analysis.  Figure ~\ref{fig:word_cloud} shows a word cloud of participants' responses.  The most frequently used terms were ``methodology'', ``data'', and ``data model'',  followed by ``know'' and ``example''.

\begin{figure}[htbp]
    \includegraphics[width=0.9\linewidth]{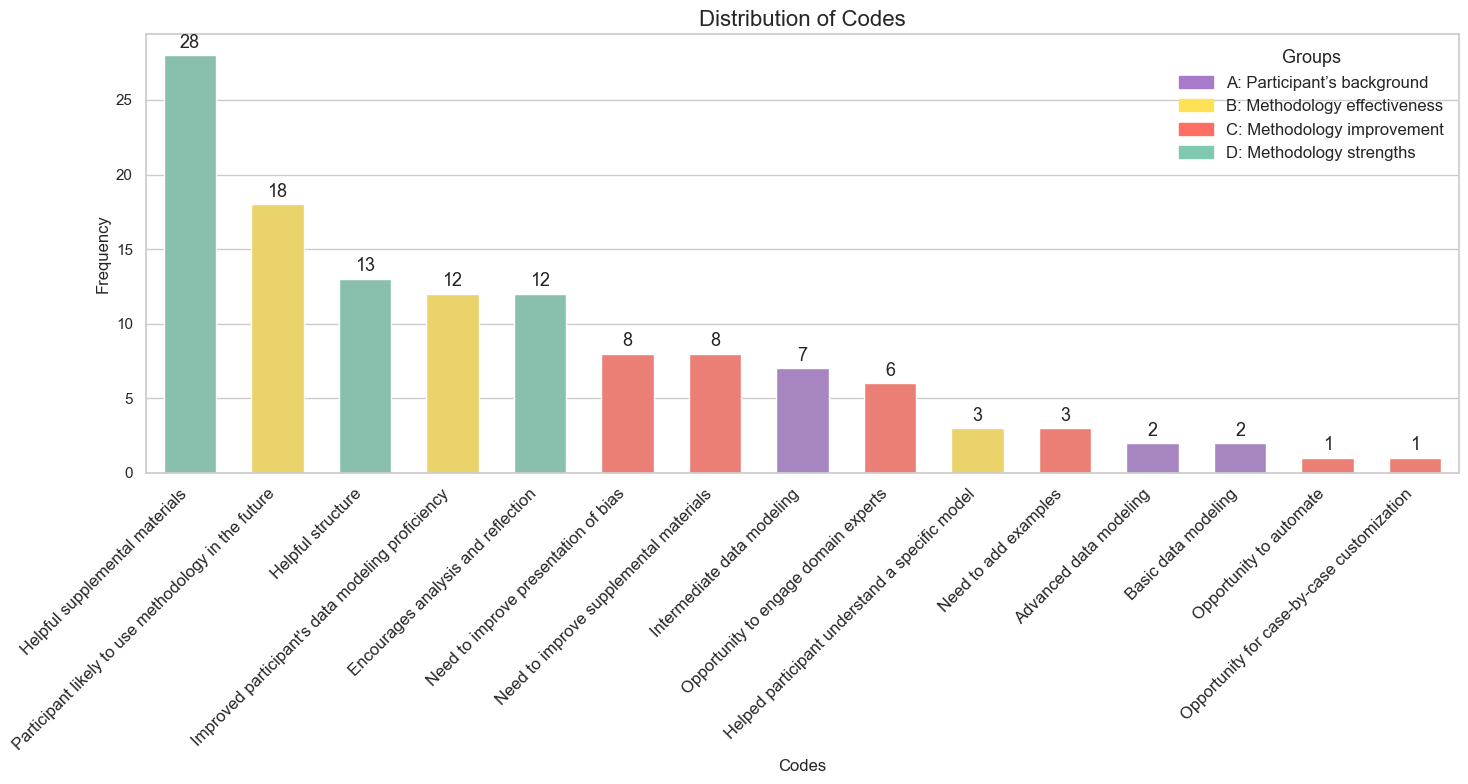}
    \caption{Interviews codes frequency distribution}
    \label{fig:code_dist}
\end{figure}

Figure~\ref{fig:code_dist} presents a summary view of the frequencies of all codes from the codebook discussed in Section~\ref{sec:eval:coding}, with colors representing our four code groups.  We observe that methodology strengths is the most frequent code group (with ``Helpful supplemental materials'', ``Helpful [methodology] structure'', and ``Encourages analysis and reflection'' as the most frequent codes), followed by methodology effectiveness (``Participant likely to use methodology in the future'' and ``Improved participant's data modeling proficiency'' as the most frequent codes), and then by methodology improvement (``Need to improve presentation of bias'' and ``Need to improve supplemental materials'' as the most frequent codes).  Based on this summary, we conjecture that \credal was found to be helpful and effective by participants, but also that there were substantial opportunities for improvement, primarily of the supplemental materials (rather than of the methodology itself).

\subsection{Research Question 1: Usability and Usefulness of \credal}

The first research question \textbf{RQ1: Is \credal usable and useful?} comprises three subquestions that we will address by analyzing participant responses.

\subsubsection{\textbf{RQ1.1}: Is \credal easy to learn and apply?}

Participants noted that the methodology features a ``Helpful Structure'' (13 code entries) and provides an abundance of ``Helpful Supplemental Materials'' (28 entries, the most frequently cited code). These findings suggest that the methodology is both feasible and effectively supported by the examples, tips, and additional resources provided, which facilitate user implementation.  Representative quotes include:

\begin{quote}
\textbf[P10]: ``I would say that the steps were defined pretty clearly. And also, there was a pretty helpful example, which went really in-depth, making it easy to work on my schema after reading the entire example with a different schema.''
\end{quote}

\begin{quote}
\textbf[P5]: ``I read that guide tutorial that was sent to me; it was very detailed and contained much information about how to perform this close reading.''
\end{quote}

\subsubsection{\textbf{RQ1.2}: Does \credal improve data modeling proficiency?}

The code ``Improved participant's data modeling proficiency'' ranks among the top five most frequently cited codes, with 12 occurrences. This code reflects participant feedback indicating that their data modeling skills and understanding significantly improved after the workshop.  Responses varied regarding the specific ways in which their skills improved, yet many participants reported gaining new conceptual and practical insights into data modeling, even those who were already familiar with the topic. Participants shared their thoughts with remarks such as:

\begin{quote}
\textbf[P5]: ``I think that understanding changed, and I can also say that I have a new view on data models after reading that tutorial.''
\end{quote}

\begin{quote}
\textbf[P9]: ``I feel more confident; gaining new knowledge about data modeling has strengthened my skills.''
\end{quote}

\subsubsection{\textbf{RQ1.3}: Are learners likely to use CREDAL in the future?}

The code labeled ``Participant likely to use methodology in the future'' appears next, with a total of 18 occurrences. This indicates that participants are inclined to apply the methodology in future projects or other contexts, often providing specific examples of such applications. Moreover, they affirmed the potential applications of this method in various scenarios, as illustrated by their comments:

\begin{quote}
\textbf[P8]: ``Enhancing the training data, especially for R\&D projects, I guess that might be a good push and a good idea to read the work that you've accomplished.''
\end{quote}
\begin{quote}
\textbf[P4]: ``A schematic framework that could help illustrate why the data model is good or bad.''
\end{quote}

Based on these results, we conclude that \credal was found to be usable and useful by participants. 

\subsection{Research Question 2: \credal helps work with data models}

The second research question \textbf{RQ2: Does \credal help understand, design, and critically evaluate data models?} focuses on the practical application of the \credal and its interaction with real data models, leading to three subquestions.

\subsubsection{\textbf{RQ2.1}: Does CREDAL help with understanding data models?}

The code ``Helped participant understand a specific model'' played a significant role in addressing this research question. Most participants emphasized that their data modeling skills improved, noting increased confidence in their understanding of data models after engaging with CREDAL.
\begin{quote}
    \textbf[P11]: ``It refreshed my knowledge about data modeling and also taught me some new ways to work with them, especially to analyze data models more carefully and look for potential problems.``
\end{quote}
\begin{quote}
      \textbf[P7]: ``I read that tutorial and learned how to perform close reading on those data models.``
\end{quote}

\subsubsection{\textbf{RQ2.2}: Does CREDAL alter the approach to structuring and modeling data within modeling tasks?}

Participants noted that prior to the workshop, many did not consider how a model was created, which data was used, or the approach developers followed in constructing a specific data model. However, after the workshop, participants observed a shift in perspective regarding data modeling tasks, indicating a willingness to adjust their approaches to create clearer data models. This is evident in quotations coded with ``Improved participant's data modeling proficiency'': 

\begin{quote}
      \textbf[P8]: ``Definitely, I guess I’ve almost never taken care about where the data was sourced and why someone decided to model it in a certain way.``
\end{quote}
\begin{quote}
      \textbf[P11]: ``The workshop provided me with a structured way to evaluate data models. It helped me spot issues before they became problems.``
\end{quote}

\subsubsection{\textbf{RQ2.3}: Do learners experience a change in their perspectives on data modeling after working with CREDAL?}

Most participants had an intermediate level of understanding of data modeling; however, even those with prior experience discovered new insights and gained unique knowledge relevant to their education or work. The same code, ``Improved participant's data modeling proficiency', captured all quotations reflecting how the methodology influenced their skills and understanding.  Representative examples are included below:

\begin{quote}
      \textbf[P3]: ``You can identify where you might be wrong about your granularity or your assumptions about whether a data model could work here or not. Without some methodology, you might miss those crucial aspects, even if you have experience working with other models.''
\end{quote}
\begin{quote}
      \textbf[P1]: ``It's essential to ensure that biases don’t negatively impact the model's performance.''
\end{quote}

Based on the analysis of the codes and provided quotations, we conclude that participants found \credal to be helpful in their work with data models.  Participants reported improvements in their ability to analyze and evaluate data models. They also proposed new approaches for improving data model design using \credal. These factors demonstrate that \credal effectively aids learners in developing a comprehensive understanding and application of data modeling practices.

\subsection{Opportunities for Improvement and Extension}
\label{sec:improvements}
Methodology improvement and extension were an essential focus of the interviews, as participants provided valuable feedback on enhancing its clarity, presentation, and practicality. The identified areas were categorized into six distinct codes, each reflecting specific concerns or suggestions from the respondents. Below, we discuss each code and its relevance to the refinement of the methodology.

\subsubsection{Improving \credal} We first look at the codes in this group relevant to improving the methodology and materials.

\paragraph{Need to Improve Supplemental Materials}
One of the recurring themes in the interviews was the need for clearer and more structured supplemental materials. Several respondents expressed difficulties in understanding the methodology from the provided text files. As one participant mentioned:

\begin{quote}
    \textbf[P1]: ``The understanding of the methodology was a bit confusing... So like to understand how it is done from the provided file. I believe that the whole methodology can be expressed in a more concise and step-by-step approach.''  
\end{quote}

This feedback emphasizes the importance of refining the materials to make the methodology more digestible. Another respondent reinforced this point by suggesting the need to simplify the description further:

\begin{quote}
   \textbf[P1]: `` We can recap the suggestions mentioned before to simplify the description of the methodology... It can be made more concise and understandable.''
\end{quote}

Analyzing data models can be a resource-intensive task. Therefore, our objective, in addition to offering a systematic approach through the CREDAL methodology, is to streamline the learning process and ensure that the methodology can be applied as efficiently and effectively as possible.

\paragraph{Need to Add More Examples}
Respondents also highlighted the importance of including more examples, particularly diagrams or visual representations, to support the text-heavy material. One participant noted:

\begin{quote}
    \textbf[P3]: ``If we're talking about the text version, I maybe added more schemas, diagrams, something like that, because it was mostly text.''
\end{quote}

Another respondent emphasized the need for examples that could help identify more different types of biases present in a data model:

\begin{quote}
    \textbf[P10]: ``I think maybe for the making assumptions part and for detecting biases, it would be helpful to use another example to just provide more types of biases that could be present in a certain data model.''
\end{quote}

As one can conclude from these quotes, incorporating visual aids, such as diagrams, ensures that users have a more interactive and comprehensive understanding of the process. Additionally, as deduced from the feedback, integrating case-oriented studies alongside theoretical explanations enhances both the clarity and applicability of the methodology. This combination allows users to grasp the methodology more effectively.

\paragraph{Need to Improve Presentation of Bias} Bias detection was a key area where respondents felt the methodology could be clearer. Participants expressed the need for a more structured approach to identifying and addressing bias within data models. One interviewee commented:

\begin{quote} \textbf[P9]: ``There is a need for clearer bias definition.'' \end{quote}

Another respondent highlighted the inherent subjectivity involved in understanding bias, making it difficult to provide a universally applicable solution:

\begin{quote} \textbf[P11]: ``My understanding of bias can be different than someone else's. So it includes a lot of subjectivity, and it's, well, it's difficult to try to avoid that.'' \end{quote}

This feedback underscores the complexity of addressing all potential forms of bias, as bias can often be perceived differently depending on the individual's perspective. While it may not be possible to cover every aspect of bias, the methodology aims to provide the broadest possible coverage, offering examples and guidance to help users recognize and mitigate harmful biases effectively within their data models.

\subsubsection{Extending \credal}
Next, we look at the codes in this group that are relevant to extending the methodology and materials.

\paragraph{Opportunity to Engage Domain Experts} Participants noted that involving domain experts could enhance the quality of the close reading process, particularly when scaling the methodology for larger teams or projects. One participant observed:

\begin{quote} \textbf[P2]: ``Engaging domain experts might improve overall close reading results when scaling the methodology for teams.'' \end{quote}

Another participant highlighted the challenges of manually identifying similar data models (here in the form of relational database schemas), stressing the potential value of expert involvement:

\begin{quote} \textbf[P3]: ``Because to try to find similar schemas by yourself, it's pretty annoying, I would say.'' \end{quote}

These suggestions point to the potential benefits of collaboration with domain specialists for a given data model to ensure a more informed and contextually relevant application of the methodology. Domain experts can provide insights into data model complexities and nuances that may be difficult for non-experts to identify, thereby improving the precision and quality of close reading analysis. Furthermore, the methodology is conveniently scalable for use not only by individual data model developers or users but also by entire teams working collaboratively on a given data model. This scalability highlights the adaptability of the approach for both small-scale and large-scale projects.

\paragraph{Opportunity to Automate} One participant repeatedly proposed automating certain parts of the methodology to make it more efficient and reduce manual workload. This respondent suggested:

\begin{quote} \textbf[P2]: ``Doing it manually is a little bit over-engineering. I would suggest trying to automate this methodology.'' \end{quote}

The participant also provided a detailed vision of how automation could be implemented, suggesting the use of large language models or similar services to streamline schema comparisons and analysis:

\begin{quote} \textbf[P3]: ``It would be cool to have something like a service where I have data, and I just apply it, or some piece of data... The service iteratively tries to analyze it based on some rules.'' \end{quote}

While automation presents a promising direction for improving efficiency, it is important to note that fully automating the entire process could introduce new sources of bias, especially if the model's decision-making lacks transparency. However, there are specific areas within the methodology that are well-suited for automation without compromising the integrity of the analysis. For instance, tasks such as related schema exploration or certain aspects of exploratory data analysis could be streamlined through automated tools \cite{abedjan,downey}. Automating these components would reduce manual effort while maintaining the critical human oversight required to ensure the accuracy and fairness of the analysis.

These insights highlight the potential for developing an automated framework that could facilitate the application of the methodology. Such a framework would enhance the accessibility and scalability of the methodology, particularly for larger data models. This suggestion aligns with future work directions, as automating the process could significantly reduce the manual effort required and contribute to making the methodology not only more efficient but also easier and more accessible for a wide range of users.

\paragraph{Opportunity for Case-by-Case Customization} Respondents also raised the idea of customizing the methodology for different use cases, suggesting that a more flexible approach could cater to varying needs. One participant explained:

\begin{quote} \textbf[P8]: ``I would like the methodology to be split into two methodologies—a longer and shorter version—for different application cases.'' \end{quote}

This feedback highlights the potential for adapting the methodology to suit a range of applications better. Rather than maintaining a single comprehensive approach, respondents suggested offering two distinct versions of the methodology: one that provides a detailed, in-depth process for complex cases and another that offers a streamlined version for simpler scenarios.

Generalizing the CREDAL methodology to apply broadly in all cases may make it overly cumbersome for certain contexts. As such, this feedback opens the door for future work in adapting the methodology by separating it according to different aspects of the application. This could involve creating a more modular framework where users can select the version of the methodology best suited to their specific needs, ensuring both efficiency and relevance in various data modeling scenarios.

\section{Conclusion, Limitations, and Future Work}
\label{sec:conc}
In our work with and within data and data systems, close readings of data models reconnect us with the materiality, the genealogies, the techne, the closed nature, and the design of data models.  
We presented the \credal methodology for close readings of data models, along with the results of a qualitative study demonstrating its usability, usefulness, and effectiveness in the critical study of data.   \credal is the first systematic method for this important activity.

We conclude with a discussion of the limitations of this study and pointers for future work.   (1) The development of \credal and our qualitative study of the methodology was limited to undergraduate and graduate students, some with professional work experience in software engineering and data science roles.  Furthermore, our primary focus was on data models for relational data systems.  Future work to address these limitations can include larger qualitative and quantitative studies in contexts beyond university students and with non-relational data modeling paradigms such as knowledge graphs and semantic web ontologies. (2) Our interview study highlighted several limitations of the current version of \credal as well as points for extension: extending and improving the supplemental materials, support and guidance for engaging with domain experts, case-by-case customization of the methodology, and opportunities to automate repetitive and data-intensive aspects of the methodology.  Addressing each of these are important research and development topics for future iterations of the methodology, and its supporting materials and technologies.
\section{Acknowledgements}
\label{sec:ack}

This research was conducted as part of the \href{https://r-ai.co/ukraine}{RAI for Ukraine} research program, run by the Center for Responsible AI at New York University in collaboration with Ukrainian Catholic University in Lviv, Ukraine.   This research was supported in part by a grant from the Simons Foundation (SFARI award \#1280457, JS).

\bibliographystyle{ACM-Reference-Format}
\bibliography{main}


\begin{thebibliography}{50}


\ifx \showCODEN    \undefined \def \showCODEN     #1{\unskip}     \fi
\ifx \showDOI      \undefined \def \showDOI       #1{#1}\fi
\ifx \showISBNx    \undefined \def \showISBNx     #1{\unskip}     \fi
\ifx \showISBNxiii \undefined \def \showISBNxiii  #1{\unskip}     \fi
\ifx \showISSN     \undefined \def \showISSN      #1{\unskip}     \fi
\ifx \showLCCN     \undefined \def \showLCCN      #1{\unskip}     \fi
\ifx \shownote     \undefined \def \shownote      #1{#1}          \fi
\ifx \showarticletitle \undefined \def \showarticletitle #1{#1}   \fi
\ifx \showURL      \undefined \def \showURL       {\relax}        \fi
\providecommand\bibfield[2]{#2}
\providecommand\bibinfo[2]{#2}
\providecommand\natexlab[1]{#1}
\providecommand\showeprint[2][]{arXiv:#2}

\bibitem[Abedjan et~al\mbox{.}(2018)]%
        {abedjan}
\bibfield{author}{\bibinfo{person}{Ziawasch Abedjan}, \bibinfo{person}{Lukasz Golab}, \bibinfo{person}{Felix Naumann}, {and} \bibinfo{person}{Thorsten Papenbrock}.} \bibinfo{year}{2018}\natexlab{}.
\newblock \bibinfo{booktitle}{\emph{Data profiling}}.
\newblock \bibinfo{publisher}{Morgan {\&} Claypool Publishers}.
\newblock


\bibitem[Akoka et~al\mbox{.}(2024)]%
        {akoka}
\bibfield{author}{\bibinfo{person}{Jacky Akoka}, \bibinfo{person}{Isabelle Comyn-Wattiau}, \bibinfo{person}{Nicolas Prat}, {and} \bibinfo{person}{Veda~C. Storey}.} \bibinfo{year}{2024}\natexlab{}.
\newblock \showarticletitle{Unraveling the foundations and the evolution of conceptual modeling—Intellectual structure, current themes, and trajectories}.
\newblock \bibinfo{journal}{\emph{Data \& Knowledge Engineering}}  \bibinfo{volume}{154} (\bibinfo{year}{2024}), \bibinfo{pages}{102351}.
\newblock


\bibitem[Ansorge(2016)]%
        {ansorge}
\bibfield{author}{\bibinfo{person}{Josef Ansorge}.} \bibinfo{year}{2016}\natexlab{}.
\newblock \bibinfo{booktitle}{\emph{Identify and sort}}.
\newblock \bibinfo{publisher}{Hurst and Co.}
\newblock


\bibitem[Azoulay(2019)]%
        {azoulay}
\bibfield{author}{\bibinfo{person}{Ariella{~}Aïsha Azoulay}.} \bibinfo{year}{2019}\natexlab{}.
\newblock \bibinfo{booktitle}{\emph{Potential history: Unlearning imperialism}}.
\newblock \bibinfo{publisher}{Verso}.
\newblock


\bibitem[Becker(2023)]%
        {becker}
\bibfield{author}{\bibinfo{person}{Christoph Becker}.} \bibinfo{year}{2023}\natexlab{}.
\newblock \bibinfo{booktitle}{\emph{Insolvent}}.
\newblock \bibinfo{publisher}{MIT Press}.
\newblock


\bibitem[Benjamin(2019)]%
        {benjamin}
\bibfield{author}{\bibinfo{person}{Ruha Benjamin}.} \bibinfo{year}{2019}\natexlab{}.
\newblock \bibinfo{booktitle}{\emph{Race after technology}}.
\newblock \bibinfo{publisher}{Polity}.
\newblock


\bibitem[Bode and Goodlad(2023)]%
        {bode}
\bibfield{author}{\bibinfo{person}{Katherine Bode} {and} \bibinfo{person}{Lauren M.~E. Goodlad}.} \bibinfo{year}{2023}\natexlab{}.
\newblock \showarticletitle{{Data worlds: An introduction}}.
\newblock \bibinfo{journal}{\emph{Critical AI}} \bibinfo{volume}{1}, \bibinfo{number}{1-2} (\bibinfo{year}{2023}).
\newblock


\bibitem[Bopp et~al\mbox{.}(2019)]%
        {bopp}
\bibfield{author}{\bibinfo{person}{Chris Bopp}, \bibinfo{person}{Lehn~M. Benjamin}, {and} \bibinfo{person}{Amy Voida}.} \bibinfo{year}{2019}\natexlab{}.
\newblock \showarticletitle{The coerciveness of the primary key: Infrastructure problems in human services work}.
\newblock \bibinfo{journal}{\emph{Proc. ACM Hum.-Comput. Interact.}} \bibinfo{volume}{3}, \bibinfo{number}{CSCW}, Article \bibinfo{articleno}{51} (\bibinfo{year}{2019}).
\newblock


\bibitem[Bowker and Star(1999)]%
        {sorting}
\bibfield{author}{\bibinfo{person}{Geoffrey~C. Bowker} {and} \bibinfo{person}{Susan~Leigh Star}.} \bibinfo{year}{1999}\natexlab{}.
\newblock \bibinfo{booktitle}{\emph{Sorting things out}}.
\newblock \bibinfo{publisher}{MIT Press}.
\newblock


\bibitem[Costanza-Chock(2020)]%
        {costanza}
\bibfield{author}{\bibinfo{person}{Sasha Costanza-Chock}.} \bibinfo{year}{2020}\natexlab{}.
\newblock \bibinfo{booktitle}{\emph{Design justice}}.
\newblock \bibinfo{publisher}{MIT Press}.
\newblock


\bibitem[Couldry and Mejias(2019)]%
        {couldry}
\bibfield{author}{\bibinfo{person}{Nick Couldry} {and} \bibinfo{person}{Ulises~A. Mejias}.} \bibinfo{year}{2019}\natexlab{}.
\newblock \bibinfo{booktitle}{\emph{The costs of connection}}.
\newblock \bibinfo{publisher}{Stanford University Press}.
\newblock


\bibitem[Dangol and Dasgupta(2023)]%
        {dangol}
\bibfield{author}{\bibinfo{person}{Aayushi Dangol} {and} \bibinfo{person}{Sayamindu Dasgupta}.} \bibinfo{year}{2023}\natexlab{}.
\newblock \showarticletitle{Constructionist approaches to critical data literacy: A review}. In \bibinfo{booktitle}{\emph{ACM IDC}}. \bibinfo{pages}{112–123}.
\newblock


\bibitem[\d{O}n\d{u}\d{o}ha(2016)]%
        {onuoha}
\bibfield{author}{\bibinfo{person}{Mimi \d{O}n\d{u}\d{o}ha}.} \bibinfo{year}{2016}\natexlab{}.
\newblock \bibinfo{title}{On missing data sets}.
\newblock \bibinfo{howpublished}{https://github.com/MimiOnuoha/missing-datasets}.
\newblock


\bibitem[Dourish(2017)]%
        {dourish}
\bibfield{author}{\bibinfo{person}{Paul Dourish}.} \bibinfo{year}{2017}\natexlab{}.
\newblock \bibinfo{booktitle}{\emph{The stuff of bits}}.
\newblock \bibinfo{publisher}{MIT Press}.
\newblock


\bibitem[Downey(2014)]%
        {downey}
\bibfield{author}{\bibinfo{person}{Allen Downey}.} \bibinfo{year}{2014}\natexlab{}.
\newblock \bibinfo{booktitle}{\emph{Think Stats: Exploratory data analysis, 2nd Ed}}.
\newblock \bibinfo{publisher}{O'Reilly}.
\newblock


\bibitem[Eubanks(2018)]%
        {eubanks}
\bibfield{author}{\bibinfo{person}{Virginia Eubanks}.} \bibinfo{year}{2018}\natexlab{}.
\newblock \bibinfo{booktitle}{\emph{Automating inequality}}.
\newblock \bibinfo{publisher}{St Martin’s Press}.
\newblock


\bibitem[Feinberg(2017a)]%
        {Feinberg17}
\bibfield{author}{\bibinfo{person}{Melanie Feinberg}.} \bibinfo{year}{2017}\natexlab{a}.
\newblock \showarticletitle{A design perspective on data}. In \bibinfo{booktitle}{\emph{{CHI}}}. \bibinfo{publisher}{{ACM}}, \bibinfo{pages}{2952--2963}.
\newblock


\bibitem[Feinberg(2017b)]%
        {melanieSlow}
\bibfield{author}{\bibinfo{person}{Melanie Feinberg}.} \bibinfo{year}{2017}\natexlab{b}.
\newblock \showarticletitle{Reading databases: slow information interactions beyond the retrieval paradigm}.
\newblock \bibinfo{journal}{\emph{Journal of Documentation}} \bibinfo{volume}{73}, \bibinfo{number}{2} (\bibinfo{year}{2017}), \bibinfo{pages}{336--356}.
\newblock


\bibitem[Feinberg(2022)]%
        {feinberg}
\bibfield{author}{\bibinfo{person}{Melanie Feinberg}.} \bibinfo{year}{2022}\natexlab{}.
\newblock \bibinfo{booktitle}{\emph{Everday adventures with unruly data}}.
\newblock \bibinfo{publisher}{MIT Press}.
\newblock


\bibitem[Foucault(1980)]%
        {foucault}
\bibfield{author}{\bibinfo{person}{Michel Foucault}.} \bibinfo{year}{1980}\natexlab{}.
\newblock \bibinfo{booktitle}{\emph{Power/Knowledge: Selected interviews and other writings 1972-1977, C. Gordon (Ed.)}}.
\newblock \bibinfo{publisher}{Pantheon Books}.
\newblock


\bibitem[Hermans and Schlesinger(2024)]%
        {hermans}
\bibfield{author}{\bibinfo{person}{Felienne Hermans} {and} \bibinfo{person}{Ari Schlesinger}.} \bibinfo{year}{2024}\natexlab{}.
\newblock \showarticletitle{A case for feminism in programming language design}. In \bibinfo{booktitle}{\emph{ACM SIGPLAN Onward!}} \bibinfo{pages}{205–222}.
\newblock


\bibitem[Hogan(2024)]%
        {hogan}
\bibfield{author}{\bibinfo{person}{Mél Hogan}.} \bibinfo{year}{2024}\natexlab{}.
\newblock \showarticletitle{{The fumes of AI}}.
\newblock \bibinfo{journal}{\emph{Critical AI}} \bibinfo{volume}{2}, \bibinfo{number}{1} (\bibinfo{date}{04} \bibinfo{year}{2024}).
\newblock


\bibitem[Kinnee et~al\mbox{.}(2023)]%
        {kinnee}
\bibfield{author}{\bibinfo{person}{Brian Kinnee}, \bibinfo{person}{Audrey Desjardins}, {and} \bibinfo{person}{Daniela Rosner}.} \bibinfo{year}{2023}\natexlab{}.
\newblock \showarticletitle{Autospeculation: Reflecting on the intimate and imaginative capacities of data analysis}. In \bibinfo{booktitle}{\emph{ACM CHI}}. Article \bibinfo{articleno}{151}, \bibinfo{numpages}{10}~pages.
\newblock


\bibitem[Lentricchia and DuBois(2003)]%
        {frank}
\bibfield{editor}{\bibinfo{person}{Frank Lentricchia} {and} \bibinfo{person}{Andrew DuBois}} (Eds.). \bibinfo{year}{2003}\natexlab{}.
\newblock \bibinfo{booktitle}{\emph{Close reading: The reader}}.
\newblock \bibinfo{publisher}{Duke UP}.
\newblock


\bibitem[Leonelli(2015)]%
        {Leonelli2015}
\bibfield{author}{\bibinfo{person}{Sabina Leonelli}.} \bibinfo{year}{2015}\natexlab{}.
\newblock \showarticletitle{What counts as scientific data? A relational framework}.
\newblock \bibinfo{journal}{\emph{Philosophy of Science}} \bibinfo{volume}{82}, \bibinfo{number}{5} (\bibinfo{year}{2015}), \bibinfo{pages}{810--821}.
\newblock


\bibitem[Leonelli and Tempini(2020)]%
        {datajourney}
\bibfield{editor}{\bibinfo{person}{Sabina Leonelli} {and} \bibinfo{person}{Niccolò Tempini}} (Eds.). \bibinfo{year}{2020}\natexlab{}.
\newblock \bibinfo{booktitle}{\emph{Data journeys in the sciences}}.
\newblock \bibinfo{publisher}{Springer}.
\newblock


\bibitem[Lukyanenko et~al\mbox{.}(2023)]%
        {LukyanenkoBSP023}
\bibfield{author}{\bibinfo{person}{Roman Lukyanenko}, \bibinfo{person}{Dominik Bork}, \bibinfo{person}{Veda~C. Storey}, \bibinfo{person}{Jeffrey Parsons}, {and} \bibinfo{person}{Oscar Pastor}.} \bibinfo{year}{2023}\natexlab{}.
\newblock \showarticletitle{Inclusive conceptual modeling: Diversity, equity, involvement, and belonging in conceptual modeling}. In \bibinfo{booktitle}{\emph{{ER} Forum}}.
\newblock


\bibitem[Malevé(2021)]%
        {maleve}
\bibfield{author}{\bibinfo{person}{Nicolas Malevé}.} \bibinfo{year}{2021}\natexlab{}.
\newblock \showarticletitle{On the data set’s ruins}.
\newblock \bibinfo{journal}{\emph{AI and Society}}  \bibinfo{volume}{36} (\bibinfo{year}{2021}), \bibinfo{pages}{1117–1131}.
\newblock


\bibitem[Martin and Taylor(2021)]%
        {martin}
\bibfield{author}{\bibinfo{person}{Aaron Martin} {and} \bibinfo{person}{Linnet Taylor}.} \bibinfo{year}{2021}\natexlab{}.
\newblock \showarticletitle{Exclusion and inclusion in identification: regulation, displacement and data justice}.
\newblock \bibinfo{journal}{\emph{Information Technology for Development}} \bibinfo{volume}{27}, \bibinfo{number}{1} (\bibinfo{year}{2021}), \bibinfo{pages}{50--66}.
\newblock


\bibitem[Muller et~al\mbox{.}(2019)]%
        {muller}
\bibfield{author}{\bibinfo{person}{Michael Muller}, \bibinfo{person}{Ingrid Lange}, \bibinfo{person}{Dakuo Wang}, \bibinfo{person}{David Piorkowski}, \bibinfo{person}{Jason Tsay}, \bibinfo{person}{Q.~Vera Liao}, \bibinfo{person}{Casey Dugan}, {and} \bibinfo{person}{Thomas Erickson}.} \bibinfo{year}{2019}\natexlab{}.
\newblock \showarticletitle{How data science workers work with data: Discovery, capture, curation, design, creation}. In \bibinfo{booktitle}{\emph{ACM CHI}}. \bibinfo{pages}{1–15}.
\newblock


\bibitem[Oluwadamilola et~al\mbox{.}(2017)]%
        {erd}
\bibfield{author}{\bibinfo{person}{Oyetola Oluwadamilola}, \bibinfo{person}{Ayodeji Okubanjo}, {and} \bibinfo{person}{Olawale Olaluwoye}.} \bibinfo{year}{2017}\natexlab{}.
\newblock \showarticletitle{A Secure Students Attendance Monitoring System}.
\newblock \bibinfo{journal}{\emph{Journal of Engineering and Technology}} \bibinfo{volume}{2}, \bibinfo{number}{2} (\bibinfo{year}{2017}).
\newblock


\bibitem[Parrish(2016)]%
        {parrish}
\bibfield{author}{\bibinfo{person}{Allison Parrish}.} \bibinfo{year}{2016}\natexlab{}.
\newblock \bibinfo{title}{Programming is forgetting}.
\newblock \bibinfo{howpublished}{http://opentranscripts.org/transcript/programming-forgetting-new-hacker-ethic/}.
\newblock
\newblock
\shownote{Keynote, Open Hardware Summit 2016}.


\bibitem[Penn(2023)]%
        {penn}
\bibfield{author}{\bibinfo{person}{Jonnie Penn}.} \bibinfo{year}{2023}\natexlab{}.
\newblock \showarticletitle{Animo Nullius: On AI’s origin story and a data colonial doctrine of discovery}.
\newblock \bibinfo{journal}{\emph{BJHS Themes}}  \bibinfo{volume}{8} (\bibinfo{year}{2023}), \bibinfo{pages}{19--34}.
\newblock


\bibitem[Poirier(2021)]%
        {poirier}
\bibfield{author}{\bibinfo{person}{Lindsay Poirier}.} \bibinfo{year}{2021}\natexlab{}.
\newblock \showarticletitle{Reading datasets: Strategies for interpreting the politics of data signification}.
\newblock \bibinfo{journal}{\emph{Big Data \& Society}} \bibinfo{volume}{8}, \bibinfo{number}{2} (\bibinfo{year}{2021}).
\newblock


\bibitem[Randell-Moon and Tippet(2016)]%
        {biopower}
\bibfield{editor}{\bibinfo{person}{Holly Randell-Moon} {and} \bibinfo{person}{Ryan Tippet}} (Eds.). \bibinfo{year}{2016}\natexlab{}.
\newblock \bibinfo{booktitle}{\emph{Security, race, biopower: Essays on technology and corporeality}}.
\newblock \bibinfo{publisher}{Palgrave Macmillan}.
\newblock


\bibitem[Sambasivan et~al\mbox{.}(2021)]%
        {SambasivanKHAPA21}
\bibfield{author}{\bibinfo{person}{Nithya Sambasivan}, \bibinfo{person}{Shivani Kapania}, \bibinfo{person}{Hannah Highfill}, \bibinfo{person}{Diana Akrong}, \bibinfo{person}{Praveen~K. Paritosh}, {and} \bibinfo{person}{Lora Aroyo}.} \bibinfo{year}{2021}\natexlab{}.
\newblock \showarticletitle{"Everyone wants to do the model work, not the data work": Data cascades in high-stakes {AI}}. In \bibinfo{booktitle}{\emph{{CHI}}}. \bibinfo{pages}{39:1--39:15}.
\newblock


\bibitem[Shaw(2023)]%
        {shaw}
\bibfield{author}{\bibinfo{person}{Ryan Shaw}.} \bibinfo{year}{2023}\natexlab{}.
\newblock \showarticletitle{Conceptual modeling as language design}.
\newblock \bibinfo{journal}{\emph{J Assoc Inf Sci Technol}} (\bibinfo{year}{2023}).
\newblock


\bibitem[Sherman et~al\mbox{.}(2024)]%
        {sherman}
\bibfield{author}{\bibinfo{person}{Jihan Sherman}, \bibinfo{person}{Romi Morrison}, \bibinfo{person}{Lauren Klein}, {and} \bibinfo{person}{Daniela Rosner}.} \bibinfo{year}{2024}\natexlab{}.
\newblock \showarticletitle{The power of absence: Thinking with archival theory in algorithmic design}. In \bibinfo{booktitle}{\emph{ACM DIS}}. \bibinfo{pages}{214–223}.
\newblock


\bibitem[Silberschatz et~al\mbox{.}(2019)]%
        {silberschatz}
\bibfield{author}{\bibinfo{person}{Avi Silberschatz}, \bibinfo{person}{Henry~F. Korth}, {and} \bibinfo{person}{S. Sudarshan}.} \bibinfo{year}{2019}\natexlab{}.
\newblock \bibinfo{booktitle}{\emph{Database System Concepts, Seventh Edition}}.
\newblock \bibinfo{publisher}{McGraw-Hill}.
\newblock


\bibitem[Simsion et~al\mbox{.}(2012)]%
        {simsion}
\bibfield{author}{\bibinfo{person}{Graeme Simsion}, \bibinfo{person}{Simon~K. Milton}, {and} \bibinfo{person}{Graeme Shanks}.} \bibinfo{year}{2012}\natexlab{}.
\newblock \showarticletitle{Data modeling: Description or design?}
\newblock \bibinfo{journal}{\emph{Information \& Management}} \bibinfo{volume}{49}, \bibinfo{number}{3} (\bibinfo{year}{2012}), \bibinfo{pages}{151--163}.
\newblock


\bibitem[Smith(1985)]%
        {Smith85}
\bibfield{author}{\bibinfo{person}{Brian~Cantwell Smith}.} \bibinfo{year}{1985}\natexlab{}.
\newblock \showarticletitle{The limits of correctness}.
\newblock \bibinfo{journal}{\emph{{SIGCAS} Comput. Soc.}} \bibinfo{volume}{14-15}, \bibinfo{number}{4, 1-3} (\bibinfo{year}{1985}), \bibinfo{pages}{18--26}.
\newblock


\bibitem[Stevens(2022)]%
        {stevens}
\bibfield{author}{\bibinfo{person}{Nikki~Lane Stevens}.} \bibinfo{year}{2022}\natexlab{}.
\newblock \emph{\bibinfo{title}{Modeling power: data models and the production of social inequality}}.
\newblock \bibinfo{thesistype}{Ph.\,D. Dissertation}. \bibinfo{school}{Arizona State University}.
\newblock


\bibitem[Storey et~al\mbox{.}(2023)]%
        {storey}
\bibfield{author}{\bibinfo{person}{Veda~C. Storey}, \bibinfo{person}{Roman Lukyanenko}, {and} \bibinfo{person}{Arturo Castellanos}.} \bibinfo{year}{2023}\natexlab{}.
\newblock \showarticletitle{Conceptual modeling: Topics, themes, and technology trends}.
\newblock \bibinfo{journal}{\emph{ACM Comput. Surv.}}  \bibinfo{volume}{55}, Article \bibinfo{articleno}{317} (\bibinfo{year}{2023}).
\newblock


\bibitem[Valdivia and Tazzioli(2023)]%
        {valdivia}
\bibfield{author}{\bibinfo{person}{Ana Valdivia} {and} \bibinfo{person}{Martina Tazzioli}.} \bibinfo{year}{2023}\natexlab{}.
\newblock \showarticletitle{Datafication genealogies beyond algorithmic fairness: Making up racialised subjects}. In \bibinfo{booktitle}{\emph{ACM FAccT}}. \bibinfo{pages}{840–850}.
\newblock


\bibitem[von{~}Lewinski et~al\mbox{.}(2024)]%
        {LewinskiBS24}
\bibfield{author}{\bibinfo{person}{Kai von{~}Lewinski}, \bibinfo{person}{Michael Beurskens}, {and} \bibinfo{person}{Stefanie Scherzinger}.} \bibinfo{year}{2024}\natexlab{}.
\newblock \showarticletitle{Data modelling as a means of power: At the legal and computer science crossroads}.
\newblock \bibinfo{journal}{\emph{Comput. Law Secur. Rev.}}  \bibinfo{volume}{52} (\bibinfo{year}{2024}), \bibinfo{pages}{105865}.
\newblock


\bibitem[Weizenbaum(1976)]%
        {weizenbaum}
\bibfield{author}{\bibinfo{person}{Joseph Weizenbaum}.} \bibinfo{year}{1976}\natexlab{}.
\newblock \bibinfo{booktitle}{\emph{Computer power and human reason}}.
\newblock \bibinfo{publisher}{W. H. Freeman and Co.}
\newblock


\bibitem[Whittaker(2023)]%
        {whittaker}
\bibfield{author}{\bibinfo{person}{Meredith Whittaker}.} \bibinfo{year}{2023}\natexlab{}.
\newblock \showarticletitle{Origin stories: plantations, computers, and industrial control}.
\newblock \bibinfo{journal}{\emph{Logic(s)}}  \bibinfo{volume}{19} (\bibinfo{year}{2023}).
\newblock


\bibitem[Wickett(2023)]%
        {wickett}
\bibfield{author}{\bibinfo{person}{Karen~M. Wickett}.} \bibinfo{year}{2023}\natexlab{}.
\newblock \showarticletitle{Critical data modeling and the basic representation model}.
\newblock \bibinfo{journal}{\emph{J Assoc Inf Sci Technol}} (\bibinfo{year}{2023}).
\newblock


\bibitem[Wiggins and Jones(2023)]%
        {wiggins}
\bibfield{author}{\bibinfo{person}{Chris Wiggins} {and} \bibinfo{person}{Matthew~L. Jones}.} \bibinfo{year}{2023}\natexlab{}.
\newblock \bibinfo{booktitle}{\emph{How data happened}}.
\newblock \bibinfo{publisher}{W. W. Norton and Co.}
\newblock


\bibitem[Yun et~al\mbox{.}(2021)]%
        {yun2021}
\bibfield{author}{\bibinfo{person}{Wei Yun}, \bibinfo{person}{Xuan Zhang}, \bibinfo{person}{Zhudong Li}, \bibinfo{person}{Hui Liu}, {and} \bibinfo{person}{Mengting Han}.} \bibinfo{year}{2021}\natexlab{}.
\newblock \showarticletitle{Knowledge modeling: A survey of processes and techniques}.
\newblock \bibinfo{journal}{\emph{Int. J. Intell. Syst.}} \bibinfo{volume}{36}, \bibinfo{number}{4} (\bibinfo{year}{2021}), \bibinfo{pages}{1686–1720}.
\newblock


\end{thebibliography}

\newpage 
\appendix
\section{Interview questions}
\label{sec:questions}

In this section, we list all interview questions, categorized into four themes, as described in Section~\ref{sec:eval}.

\paragraph{Background and experience with the \credal methodology}
    \begin{itemize}
        \item [1.] Please describe your familiarity and experience with data modeling before the workshop.
        \item [2.] Did your understanding of data modeling change as a result of the workshop?  Do you feel more or less confident in your data modeling skills after the workshop? Please explain.
        \item [3.] Follow-up: Please describe your overall experience in learning and applying the \credal methodology during the workshop. 
        \item [4.] Follow-up: Which aspects of the methodology were particularly helpful? Which aspects of the methodology were particularly challenging?
    \end{itemize}

\paragraph{Feedback on supplemental materials}
    \begin{itemize}
        \item [5.] On a scale of 1 to 5, how helpful was the tutorial for understanding the methodology? (1- Not helpful, 5- Very helpful)
        \item [6.] How helpful was the example of close reading, we provided in helping for your understanding of the methodology? (1- Not helpful, 5- Very helpful)
        \item [7.] Are you familiar with the literary close reading technique? (Yes/No)
        \item [8.] If the answer was ``Yes'', did you find the comparison in the tutorial helpful for understanding our methodology?
    \end{itemize}
    
\paragraph{Perceived effectiveness of \credal}
    \begin{itemize}
        \item [9.] Were you able to apply the methodology during the reading session? (Yes/No)
        \item [10.] If the answer was ``No'', what challenges did you face?
        \item [11.] Please give an example of how you might apply \credal in your work or studies.
        \item [12.] Would you consider using the \credal methodology for ``reading of data models'' as a preliminary step for future projects? (Yes/No)
        \item [13.] If the answer was ``Yes'', at what point during your project would you use the methodology?  Will you use it as a checker before starting work with some sort of datasets or data models?
    \end{itemize}

\paragraph{Suggestions for improving \credal}
    \begin{itemize}
        \item [14.] Please suggest ways to improve the methodology itself or how we explain it.
        
    \end{itemize}

\section{Data model for example reading}
\label{sec:example_reading}

\begin{figure}[H]
    \centering
    \includegraphics[width=1\linewidth]{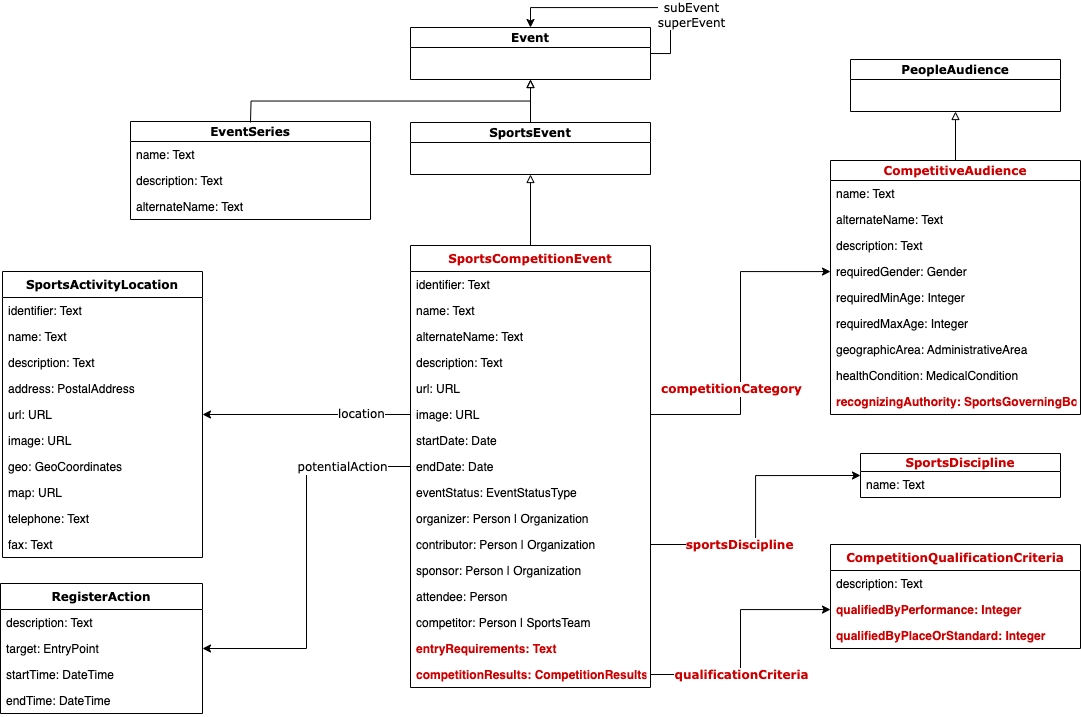}
    \caption{Data Model for Example Reading, based on~\citet{erd}}
    \label{fig:enter-label}
\end{figure}
\newpage

\section{Intreview codebook}
\label{sec:Codebook}


\begin{table}[htbp]
\centering
\small
\begin{tabular}{|>{\raggedright\arraybackslash}p{3cm}|>{\raggedright\arraybackslash}p{4cm}|>{\raggedright\arraybackslash}p{7cm}|}
\hline
\textbf{Groups} & \textbf{Codes} & \textbf{Explanation} \\ \hline

\cellcolor{blue!10}{A: Participant's background} 
& \cellcolor{blue!10}1: Basic data modeling & \cellcolor{blue!10}General feedback on respondent's previous data model experience \\ \cline{2-3}
\cellcolor{blue!10} & \cellcolor{blue!10}2: Intermediate data modeling & \cellcolor{blue!10}General feedback on respondent's previous data model experience \\ \cline{2-3}
\cellcolor{blue!10} & \cellcolor{blue!10}3: Advanced data modeling & \cellcolor{blue!10}General feedback on respondent's previous data model experience \\ \hline

\cellcolor{yellow!10}{B: Methodology effectiveness} & \cellcolor{yellow!10}1: Improved participant's data modeling proficiency& \cellcolor{yellow!10}Anytime respondents mentioned that methodology improved their data modeling skills \\ \cline{2-3}
\cellcolor{yellow!10} & \cellcolor{yellow!10}2: Helped participant understand a specific model & \cellcolor{yellow!10}Anytime respondents mentioned that methodology helped them understand a data model \\ \cline{2-3}
\cellcolor{yellow!10} & \cellcolor{yellow!10}3: Participant likely to use the methodology in the future & \cellcolor{yellow!10}Whenever respondents claimed they would use the methodology for future work, studies, or personal projects \\ \hline

\cellcolor{red!10}{C: Methodology improvement} & \cellcolor{red!10}1: Opportunity to engage domain experts & \cellcolor{red!10}When scaling methodology for teams, domain expert engagement might improve overall close reading results \\ \cline{2-3}
\cellcolor{red!10}& \cellcolor{red!10}2: Opportunity to automate & \cellcolor{red!10}When respondents identified opportunities to automate or semi-automate application of the methodology \\ \cline{2-3} 
\cellcolor{red!10} & \cellcolor{red!10}3: Need to add examples& \cellcolor{red!10}When respondents mentioned the need for more examples (in video or the example close reading) \\ \cline{2-3} 
\cellcolor{red!10} & \cellcolor{red!10}4: Need to improve supplemental materials & \cellcolor{red!10}When a responded states a need to improve supplemental materials, e.g., tutorial, examples, etc) \\ \cline{2-3}
\cellcolor{red!10} & \cellcolor{red!10}5: Need to improve presentation of bias & \cellcolor{red!10}Whenever respondents mentioned the need for clearer bias definition \\ \cline{2-3} 
\cellcolor{red!10} & \cellcolor{red!10}6: Opportunity for case-by-case customization & \cellcolor{red!10}When respondents asked for shorter and longer versions of the methodology for different application cases \\ \hline

\cellcolor{green!10}{D: Methodology strengths} 
& \cellcolor{green!10}1: Helpful structure & \cellcolor{green!10}Respondents mentioning they liked the steps in the methodology (e.g., assumption loop), as well as its general structure: divide and conquer, and iterative deliberation \\ \cline{2-3}
\cellcolor{green!10} & \cellcolor{green!10}2: Helpful supplemental materials & \cellcolor{green!10}Respondents liked the video, tips \& tricks, and example, which were in addition to the methodology \\ \cline{2-3}
\cellcolor{green!10} & \cellcolor{green!10}3: Encourages analysis and reflection & \cellcolor{green!10}Respondents mentioned that the methodology is time-consuming or requires deep data model analysis (which is what we aim for) \\ \hline

\end{tabular}
\caption{Groups, Codes, and Explanation}
\end{table} 

\end{document}